\documentclass[a4paper, accepted=2021-12-03]{quantumarticle}
 
\pdfoutput=1

\usepackage[colorlinks = true]{hyperref}
\usepackage{amsmath,amssymb,amsthm,mathbbol,bbm,bm,dsfont,xcolor,booktabs,enumitem,braket,graphicx}
\usepackage[numbers,sort&compress]{natbib}

\def\Tr{{\rm{Tr}}}
\def\croot{{\rm{i}}}
\def\econ{{\rm{e}}}

\begin{document}
\title{Relaxation times do not capture logical qubit dynamics}

\author{Amit Kumar Pal}
\affiliation{Department of Physics, Indian Institute of Technology Palakkad, Palakkad 678557, India}
\affiliation{Department of Physics, College of Science, Swansea University, Singleton Park, Swansea - SA2 8PP, United Kingdom}
\affiliation{Faculty of Physics, University of Warsaw, ul. Pasteura 5, PL-02-093 Warszawa, Poland}

\author{Philipp Schindler}
\affiliation{Institut f\"ur Experimentalphysik, Universit\"at Innsbruck, Technikerstr.  25, A-6020 Innsbruck, Austria}

\author{Alexander Erhard}
\affiliation{Institut f\"ur Experimentalphysik, Universit\"at Innsbruck, Technikerstr.  25, A-6020 Innsbruck, Austria}

\author{\'{A}ngel Rivas}
\affiliation{Departamento de F\'isica Te\'orica, Facultad de Ciencias F\'isicas,
Universidad Complutense, 28040 Madrid, Spain}
\affiliation{CCS -Center for Computational Simulation, Campus de Montegancedo UPM, 28660 Boadilla del Monte, Madrid, Spain.}

\author{Miguel-Angel Martin-Delgado}
\affiliation{Departamento de F\'isica Te\'orica, Facultad de Ciencias F\'isicas,
Universidad Complutense, 28040 Madrid, Spain}
\affiliation{CCS -Center for Computational Simulation, Campus de Montegancedo UPM, 28660 Boadilla del Monte, Madrid, Spain.}

\author{Rainer Blatt}
\affiliation{Institut f\"ur Experimentalphysik, Universit\"at Innsbruck, Technikerstr.  25, A-6020 Innsbruck, Austria}
\affiliation{Institut  f\"ur  Quantenoptik  und  Quanteninformation, \"Osterreichische  Akademie  der  Wissenschaften, Otto-Hittmair-Platz  1, A-6020 Innsbruck, Austria}

\author{Thomas Monz}
\affiliation{Institut f\"ur Experimentalphysik, Universit\"at Innsbruck, Technikerstr.  25, A-6020 Innsbruck, Austria}
\affiliation{Alpine Quantum Technologies GmbH, 6020 Innsbruck, Austria}

\author{Markus M\"uller}
\affiliation{Institute for Quantum Information, RWTH Aachen University, D-52056 Aachen, Germany}
\affiliation{Peter Gr\"{u}nberg Institute, Theoretical Nanoelectronics, Forschungszentrum J\"{u}lich, D-52425 J\"{u}lich, Germany}
\affiliation{Department of Physics, College of Science, Swansea University, Singleton Park, Swansea - SA2 8PP, United Kingdom}

\begin{abstract}
 Quantum error correction procedures have the potential to enable faithful operation of large-scale quantum computers. They protect information from environmental decoherence by storing it in logical qubits, built from ensembles of entangled physical qubits according to suitably tailored quantum error correcting encodings. To date, no generally accepted framework to characterise the behaviour of logical qubits as quantum memories has been developed. In this work, we show that generalisations of well-established figures of merit of physical qubits, such as relaxation times, to logical qubits fail and do not capture dynamics of logical qubits. We experimentally illustrate that, in particular, spatial noise correlations can give rise to rich and counter-intuitive dynamical behavior of logical qubits. We show that a suitable set of observables, formed by code space population and logical operators within the code space, allows one to track and characterize the dynamical behaviour of logical qubits. Awareness of these effects and the efficient characterisation tools used in this work will help to guide and benchmark experimental implementations of logical qubits.
\end{abstract}
 
\maketitle

\section{Introduction}
\label{sec:intro}
High-quality physical qubits with long coherence times that allow one to reliably store fragile quantum states form the backbone of currently developed quantum processors~\cite{Ladd2010,nielsen2010}. Over the last decades, the development of methods to characterise physical qubits and their coherence properties has been subject of intense study.
Here, widespread and popular figures of merit are the longitudinal and transverse relaxation time scales, known as $T_1$ and $T_2$. They were originally introduced in the field of nuclear magnetic resonance, describing a simple exponential decay dynamics of spin states \cite{nielsen2010,abragam1961}.

Such simple descriptions, however, become incomplete in the presence of, e.g., temporal noise correlations giving rise to non-Markovian dynamics~\cite{breuer-book,rivasRMP-2014}. Similarly, spatial noise correlations can play a role in larger quantum registers, where such correlations can be quantified and measured \cite{Postler2018,Rivas-Mueller-2015} and sometimes also harnessed for noise mitigation techniques, for instance by storing quantum information in decoherence-free subspaces \cite{zanardi1997,lidar1998,lidar2001a,lidar2001b,kielpinski1013,haffner2005}.

Currently, we are witnessing enormous efforts to build and reliably control increasingly larger quantum processors - often termed noisy intermediate-scale quantum (NISQ) devices \cite{Preskill2018}. These devices are also used to implement low-distance quantum error correcting codes \cite{Chiaverini2004,Schindler2011,Reed2012,Nigg2014,Waldherr2014,kelly2015,Linke2017,arXiv:1912.09410}, which allow one to encode and protect quantum information in so-called logical qubits formed of entangled ensembles of physical qubits \cite{Terhal-2015,lidar2013,nielsen2010}. An important short-term goal is to reduce the effective error rates \cite{Gambetta2017,bermudez-prx2017,debroy2019}, as a first step towards the long-term goal of protected large-scale fault-tolerant quantum computation \cite{kitaev2001,ShorThreshold,PreskillThreshold}.

However, characterising the performance of logical qubits is naturally more involved, because fully characterising the state of its constituents is not feasible for even intermediate-size quantum registers. 
It is tempting to try to directly leverage the well-established figures of merit developed for physical qubits to logical qubits, guided by the intuition that the encoded information in logical qubits should show qualitatively similar dynamical behaviour as their physical constituents. In this work we illustrate that the analogy to a physical qubit does not hold generally, and that the characterisation of logical qubits as quantum memories \cite{Terhal-2015} comes with a number of unique challenges. In particular, spatial noise correlations can strongly affect QEC performance \cite{clemens2004,klesse2005,aharonov2006,preskill2013,novais2013,novais2006,shabani2008} and influence dynamical behaviour of logical qubits in a counter-intuitive way.

For example, we show that generalisations of, e.g. $T_1$ and $T_2$ times to logical qubits fail, even for encodings consisting of no more than 3 or 4 physical qubits. We theoretically discuss and experimentally observe rich decay dynamics of small-scale logical qubits, due to leakage of quantum information from the code space, or temporal behavior governed by multiple time scales in contrast to simple exponential decay. We foresee that awareness of these effects and the efficient characterisation tools used in this work will guide the development and optimisation of logical qubits.

\section{Experimental system and noise}
\label{sec:experiment}
The experimental setup consists of a trapped-ion quantum information processor with $^{40}$Ca$^+$ ions, that has been described in detail in reference~\cite{Schindler2013}. The qubits are encoded in the 4S${_{1/2}(m_j=-1/2)=\ket{1}}$ ground state and the metastable excited state 3D${_{5/2}(m_j=-1/2)=\ket{0}}$  and transitions between these states are driven with a narrow linewidth laser~\cite{Schindler2013}.  The system provides a universal set of gate operations consisting of M{\o}lmer-S{\o}rensen (MS) entangling gates and arbitrary local operations
~\cite{Schindler2013,Martinez2016}. Any local operation can be implemented by a combination of a resonant collective local operation $U_{x}(\theta) = \exp(-i \theta/2 S_x)$, with $S_{x} = \sum_i X_i$ being the sum over all single-qubit $X$ Pauli operators\footnote{We denote the Pauli operators with their capital letters $X,Y,Z$ to facilitate the notation.}, and single-qubit AC-Stark shifts, represented by rotations around the z-axis of the Bloch sphere $U_z^{(i)}(\theta) = \exp(-i  \theta/2 Z_i)$.
The action of the entangling MS gate operation on the entire qubit register is described as $\mathrm{MS}(\theta)=\exp(-i  \theta/4 S_x^2)$.

The dominating noise source for storing information in our experimental system is given by dephasing caused by laser frequency noise and fluctuations in the bias magnetic field~\cite{Schindler2013}.
In our system, the effect of fluctuations of the  laser frequency as well as the magnitude of the magnetic field cannot be distinguished. We can thus describe the dephasing process using a single fluctuating variable $B(t)$, referred to in the following as effective magnetic field:
\begin{eqnarray}
H_G(t)=\frac{1}{2}B(t) Z. 
\label{eq:single_dephasing_hamiltonian}
\end{eqnarray}
In the following, we assume the random fluctuation in the values of the effective magnetic field to obey a Gaussian distribution $P(B)$, which implies that
\begin{eqnarray}
&&\left\langle\exp\left[\pm\text{i}\int_{0}^tB(t^\prime)dt^\prime\right]\right\rangle\nonumber\\
&=&\exp\left[-\frac{1}{2}\left\langle\left(\int_{0}^tB(t^\prime)dt^\prime\right)^2\right\rangle\right].
\label{eq:because_gaussian}
\end{eqnarray} 
We also assume a stationary autocorrelation function of the noise source, implying 
\begin{eqnarray}
\langle B(t+\tau)B(t)\rangle=\langle B(\tau)B(0)\rangle, 
\end{eqnarray}
and a further $\delta$-correlation of the noise, such that 
\begin{eqnarray}
\langle B(\tau)B(0)\rangle=\langle [B(0)]^2\rangle\delta(\tau).
\end{eqnarray}
Therefore, in the case of local dephasing, this implies
\begin{eqnarray}
\langle B_k(t+\tau)B_l(t)\rangle&=&\langle [B_k(0)]^2\rangle\delta_{k,l}\delta(\tau). 
\end{eqnarray}
Using these properties, one finds
\begin{eqnarray}
\left\langle\left[\int_{0}^tB(t^\prime)dt^\prime\right]^2\right\rangle
=\langle[B(0)]^2\rangle t=\gamma t,
\label{eq:define_gamma}
\end{eqnarray}
where we define $\gamma=\langle[B(0)]^2\rangle$.

We will, for completeness, now present a brief overview of the relevant results obtained when dephasing noise is applied to a single physical qubit. 
Writing a generic pure single-qubit state in terms of the computational basis $\{\ket{0},\ket{1}\}$ as $\ket{\psi}=\cos\frac{\theta}{2}\ket{0}+\text{e}^{\text{i}\phi}\sin\frac{\theta}{2}\ket{1}$, with $\theta$ and $\phi$ being real parameters ($0\leq \theta\leq \pi$, $0\leq \phi\le 2\pi$), the dephasing noise acts on the state as $\ket{\psi^\prime}=\exp\left[-\croot\int_0^t H_G(t^\prime)dt^\prime\right]\ket{\psi}_L$, leading to
\begin{eqnarray}
\ket{\psi^\prime}=\cos\frac{\theta}{2}\ket{0}
+\exp\left[\text{i}\left(\phi+\int_{0}^tB(t^\prime)dt^\prime\right)\right]\sin\frac{\theta}{2}\ket{1},\nonumber\\
\end{eqnarray}
discarding a global phase $\exp\left[-\frac{\text{i}}{2}\int_{0}^tB(t^\prime)dt^\prime Z\right]$.

Denoting the distribution of the random values of the magnetic field by $P(B)$, the density matrix of the qubit is given by $\rho^\prime=\int \ket{\psi^\prime}\bra{\psi^\prime} P(B)dB$. Assuming $P(B)$ to be a Gaussian distribution, and using Eq.~(\ref{eq:define_gamma}), the noisy density matrix can be simplified as 
\begin{eqnarray}
\rho^\prime&=& \cos^2\frac{\theta}{2}\ket{0}\bra{0}+\sin^2\frac{\theta}{2}\ket{1}\bra{1}\nonumber\\
&&+\frac{1}{2}\econ^{-\frac{1}{2}\gamma t}\sin\theta(\text{e}^{\text{-i}\phi}\ket{0}\bra{1}+\text{e}^{\text{i}\phi}\ket{1}\bra{0}).
\end{eqnarray}
For a physical qubit represented completely by its Bloch vectors $\vec{r}=(r_x,r_y,r_z)$, where $r_x\equiv X$, $r_y\equiv Y$, and $r_z\equiv Z$, it is crucial to understand how the components of the Bloch vectors are modified under the application of the dephasing noise.  The expectation values of the components of the Bloch vector in the state $\rho^\prime$ evolve under dephasing as 
\begin{eqnarray}
\label{eq:x_physical}
\langle X\rangle&=&\text{Tr}(X\rho^\prime)=\text{e}^{-\frac{1}{2}\gamma t}\sin\theta\cos\phi, \\
\label{eq:y_physical}
\langle Y\rangle&=&\text{Tr}(Y\rho^\prime)=\text{e}^{-\frac{1}{2}\gamma t}\sin\theta\sin\phi, \\
\label{z_physical}
\langle Z\rangle&=&\text{Tr}(Z\rho^\prime)=\cos\theta.
\end{eqnarray}
This behavior is a special case of the most general qubit relaxation dynamics which is characterized by the longitudinal and transverse the relaxation time-scales $T_1$ and $T_2$, as introduced in the early nuclear magnetic resonance experiments. These relaxation times are defined as
\begin{eqnarray}
\label{eq:longitudinal}
r_z(t)&=& r_z^{\text{eq}}-\econ^{-\frac{t}{T_1}}\left[r_z^{\text{eq}}-r_z(0)\right],\\
\label{eq:transverse}
r_\perp(t)&=& r_\perp^{\text{eq}}-\econ^{-\frac{t}{T_2}}\left[r_\perp^{\text{eq}}-r_\perp(0)\right],
\end{eqnarray}
where $r_\perp=\sqrt{r_x^2+r_y^2}$, and the superscript ``eq" signifies the equilibrium time of the corresponding signal when the system has fully relaxed. Here, $T_1$ represents the typical decay time of the eigenstates of the $Z$ Pauli matrix, and $T_2$ quantifies the lifetime of quantum coherence between them. Comparing Eqs.~(\ref{eq:longitudinal})-(\ref{eq:transverse}) with Eqs.~(\ref{eq:x_physical})-(\ref{eq:y_physical}),  one obtains $T_1=\infty$, while $T_2=\frac{1}{2\gamma}$ for dephasing noise. 

In a multi-qubit system, the spatial correlation of the noise needs to be accounted for. We concentrate on two extreme cases of spatial noise correlations: (i) local dephasing noise, where each qubit has its own, independent noise source, and (ii) a global, i.e.~collective dephasing where one noise source is affecting all qubits identically.

Local dephasing would be caused by local fluctuating magnetic fields, where each of the physical qubits constituting the logical qubit experiences a different random magnetic field, and the noise Hamiltonian is given by 
\begin{eqnarray}
H_L(t)=\frac{1}{2}\sum_k B_k(t)Z_k,
\label{eq:local_dephasing_hamiltonian} 
\end{eqnarray}
where $B_k(t)$ is the time-dependent strength of the magnetic field local to the physical qubit $k$, and $Z_k$ is the $z$-component of the Pauli matrices corresponding to qubit $k$.  
On the other hand, the global dephasing noise is due to a randomly fluctuating effective magnetic field that acts on all of the physical qubits, such that the noise Hamiltonian is given by 
\begin{eqnarray}
H_G(t)=\frac{1}{2}B(t)\sum_k Z_k, 
\label{eq:global_dephasing_hamiltonian}
\end{eqnarray}
where $B(t)$ is the time-dependent strength of the global fluctuating magnetic field.  
In typical ion-trap experiments, global dephasing is dominating, as the typical length-scale of noise fields is much larger than the inter-ion distance~\cite{Schindler2011,Postler2018}. Global dephasing is also applicable to any system that uses a common local oscillator as phase reference.

In the following sections, we showcase the performance of our proposed parameters in quantifying the quality of a logical qubit at the example of dephasing noise. Naturally, a similar analysis can also be carried out  with other types of noise at play, e.g. amplitude damping noise (see appendix~\ref{sec:diff_noise}).

\section{A logical qubit under dephasing}
\label{subsec:logical_qubit_dephasing}

A logical qubit is constructed from $N$ physical qubits, and its generic pure logical state is denoted as $\ket{\psi}_L=\cos\frac{\theta}{2}\ket{0}_L+\text{e}^{\text{i}\phi}\sin\frac{\theta}{2}\ket{1}_L$. The logical basis states $\{\ket{0}_L$ and $\ket{1}_L$ are, in general, $N$-qubit entangled states. A logical qubit is defined by the set of stabilizer generators $\{S_i\}$ and the set of logical operators $\{X_L,Y_L,Z_L\}$ as
\begin{eqnarray}
X_L\ket{0}_L&=&\ket{1}_L,\,\,X_L\ket{1}_L=\ket{0}_L,\\
Z_L\ket{0}_L&=&\ket{0}_L,\,\,Z_L\ket{1}_L=-\ket{1}_L.
\end{eqnarray}
Each of these logical operators is acting on multiple physical qubits. Without any loss in generality, one can express the logical state $\ket{0}_L$ as a superposition of computational basis states of the physical qubits as $\ket{0}_L = \sum_l b_l \ket{b}_l$. 

The effect of dephasing noise on such a complex $N$-qubit state can be analyzed straightforwardly by grouping the physical basis states by their \textit{magnetization}. The magnetization of a basis state is defined as the difference between the number of spins in the ground state $\ket{0}$ with eigenvalue $+1$, denoted as $n$, and the remaining number of spins in the excited state $\ket{1}$ with eigenvalue $-1$, $N-n$. The magnetization is expressed as
\begin{eqnarray}
m = 2n - N \, .
\end{eqnarray}

Each magnetization value has the multiplicity $N_m = N!/(m! (N-m)!)$. The magnetization has $N+1$ possible values given by $m \in \{-N,-N+2,\cdots,N-2,N \}$.
The logical basis state $\ket{0}_L$ can then be written by grouping the physical basis states by their magnetization:
\begin{eqnarray}
\ket{0}_L=\sum_{m} \sum_{l=1}^{N_m} b_l^m\ket{b}_l^m \, .
\end{eqnarray}
The state $\ket{1}_L$ can also be written in a similar way.

Let us first consider the global dephasing noise represented by the noise Hamiltonian $H_G(t)$. The effect of the global dephasing noise  on a generic logical state $\ket{\psi}_L$, given by $\ket{\psi^\prime}_L=\exp\left[-\croot\int_0^t H_G(t^\prime)dt^\prime\right]\ket{\psi}_L$, is determined by the eigenvalue equation 
\begin{eqnarray}
\left[\sum_kZ_k\right]\ket{b}_l^m=m|b\rangle_l^m.
\label{eq:global_dephasing_eigenvalue}
\end{eqnarray}
Therefore, in the density matrix 
$\rho=\int\left(\ket{\psi^\prime}\bra{\psi^\prime}\right)_L P(B)dB$ of the logical qubit, the off-diagonal elements $\ket{b}_l^m\bra{b}_{l^\prime}^{m^\prime}$
have coefficients decaying with time as  
\begin{eqnarray}
c_{\Delta m}&=&\exp\left[-\frac{1}{2}\left(\frac{\Delta m}{2}\right)^2\gamma t\right],
\label{eq:global_timescales}
\end{eqnarray}
where the difference in magnetization $\Delta m=m-m^\prime$ takes integer values. Note that the time-decays of these coefficients originate solely due to the difference $\Delta m=m-m^\prime$ in the \emph{magnetization} values corresponding to different basis states $\ket{b}_l^m$.  For situations where $\Delta m=0$,  no manifestation of the global noise in the form of the time-decay of the coefficients of the density matrix can be found. The subspace of the Hilbert space of the $N$-qubit system hosting the basis states for which $\Delta m=0$, therefore, forms a \emph{decoherence-free subspace} (DFS) which is not affected by global dephasing noise.

In contrast to Eq.~(\ref{eq:global_dephasing_eigenvalue}), the effect of \emph{local dephasing noise} governed by the Hamiltonian $H_L(t)$ on the logical qubit state $\ket{\psi}_L$ is determined by the eigenvalue equation 
\begin{eqnarray}
\left[\sum_k B_k(t)Z_k\right]\ket{b}_l=\left[\sum_k\alpha_kB_k(t)\right]\ket{b}_l,
\end{eqnarray}
where the factors $\alpha_k=\pm 1$ are defined by $Z_k\ket{k}_l=\alpha_k\ket{b}_l$, i.e., whether the $k$th qubit in $\ket{b}_l$ is in $\ket{0}$ or $\ket{1}$ state. 
For uncorrelated dephasing of equal strength on the $N$ qubits this leads to a decay of the off-diagonal terms in the density matrix $\rho^\prime$ as 
\begin{eqnarray}
c_{\Delta n}&=&\exp\left[-\frac{\Delta n}{2}\gamma t\right],
\label{eq:local_timescales}
\end{eqnarray}
where $\Delta n$ is the number of positions in the basis states $\ket{b}_l$ and $\ket{b}_{l^\prime}$ where the entries differ (Hamming distance). Note that in this case the dephasing dynamics is not governed by the (differences in) \textit{magnetization} $m$, and DFS does not exist in this case.

\subsection{Assessing the quality of a logical qubit}
\label{subsec:observables}

We now discuss the relevant quantities to assess the quality and to characterise decay dynamics of a logical qubit. 
A natural choice of such quantities would be the components of the \emph{logical Bloch vector} $\vec{R}=\left(R_x,R_y,R_z\right)$, where we identify $R_{x,y,z}$ as 
\begin{align}
R_x=\langle X_L\rangle,\,R_y=\langle Y_L\rangle,\,R_z=\langle Z_L\rangle.
\label{eq:logical_op_exp_values} 
\end{align}
Here, $\langle \mathcal{O}\rangle=\Tr\left[\mathcal{O}\rho^\prime\right]$ is the expectation value of the operator $\mathcal{O}$ in the noisy state $\rho^\prime$ of the logical qubit.

A major issue for characterizing logical qubit dynamics is the fact that noise typically causes leakage from the code space. It is therefore useful to also quantify the code-space population, $p=\langle P_\text{c}\rangle$, where 
\begin{eqnarray}
P_{\text{c}}=\frac{1}{2^{N}}\prod_{k=1}^N(I+S_k)
\label{eq:code_space_population_op}
\end{eqnarray}
denotes the projector onto the code-space of an $N$-qubit stabilizer QEC code \cite{nielsen2010}. Here, \{$S_k\}$ is the set of stabilizer generators that define the code, and $I$ is the identity operator in the Hilbert space of the $N$ physical qubits.
Projecting on the code-space population corresponds to post-selecting on the no-error outcome if one realized a perfect syndrome measurement, i.e.~the set of generators of the code, via ancilla qubits.

Note that the code-space population and all other quantities discussed below can be evaluated from measuring the $2^N$ stabilizer elements of the code, requiring fewer measurements than full state tomography. Furthermore, the number of measurements could be reduced further using techniques proposed in the context of efficient fidelity estimation of stabilizer states~\cite{PhysRevLett.106.230501}.

In order to incorporate the effect of leakage from the code space in the expectation values of the logical operators, we also consider the quantities $\{p_x,p_y,p_z\}$, where 
\begin{eqnarray}
p_x=\langle X_L P_\text{c}\rangle,\,p_y=\langle Y_L P_\text{c}\rangle,p_z=\langle Z_L P_\text{c}\rangle.
\label{eq:operator_expt_values_in_code_space}
\end{eqnarray}
The relevant time-scales in the evolution of these quantities under global dephasing noise are given by Eq.~(\ref{eq:global_timescales}) for magnetization differences $\Delta m$. In Sec.~\ref{subsec:illustrations}, we derive the theoretical results for the time evolution of the expectation values of these quantities. We then also compare this to our experimental results.

We stress here that these quantities are defined independent of the specific noise model.
In the following subsections, we demonstrate the performance of the quantities using dephasing noise, which is the dominant noise in the experimental setup considered in this paper. However, these quantities can also be used to investigate the quality of the logical qubit under other types of noise. In appendix~\ref{sec:diff_noise} we present a comparison of simulated global dephasing versus simulated amplitude damping noise.

\subsection{Dephasing noise on small QEC codes}
\label{subsec:illustrations}
We now examine how the quantities discussed in Sec.~\ref{subsec:observables} evolve over time under global dephasing noise, for a single logical qubit in a variety of three- and four-qubit QEC codes. 

\subsubsection{A three-qubit bit-flip code} 
The first example we consider is that of a $3$-qubit QEC code, whose stabilizer operators are given by 
\begin{eqnarray} 
S_1=Y_1X_2Y_3, \,
S_2=X_1Y_2Y_3,
\end{eqnarray}
and the logical operators are 
\begin{eqnarray}
X_L &=& -Y_1Y_2Z_3,\nonumber\\
Z_L &=& X_1X_2X_3, \nonumber\\
Y_L &=&\text{i}X_L Z_L = Z_1 Z_2 Y_3
\label{eq:logical_operator_3qubit}
\end{eqnarray}
The logical basis states $\{\ket{0}_L,\ket{1}_L\}$ are given by 
\begin{eqnarray}
\ket{0}_L&=&\frac{1}{\sqrt{2}}\left(\ket{001}+\ket{110}\right),\nonumber\\
\ket{1}_L&=&\frac{1}{\sqrt{2}}\left(\ket{000}-\ket{111}\right).
\label{eq:logical_computational_basis_3qubit}
\end{eqnarray}
The motivation behind choosing this specific form of the logical basis is two-fold. Firstly, the effect of the global dephasing noise depends explicitly on the choice of the logical basis, as explained in Sec.~\ref{subsec:logical_qubit_dephasing}. Therefore, it is important to choose a set of logical basis state that clearly demonstrates the effect of the different magnetization values, which is achieved by the chosen basis. Secondly, the chosen logical basis states are easy to prepare, using only a single MS gate.

Evidently, the basis states $\ket{b}_l^m$ contributing in $\ket{\psi}_L$ have four specific values of $m$, given by $m=-3(\ket{111})$, $-1(\ket{110})$, $1(\ket{001})$, and $3(\ket{000})$. Therefore, following the discussions in Sec.~\ref{subsec:logical_qubit_dephasing}, the dynamics of the coefficients of the off-diagonal elements in $\rho^\prime$ are governed by the exponential decay factors as given by Eq.~(\ref{eq:global_timescales}), namely $\exp\left[-\frac{1}{2}\gamma t\right]$ (corresponding to the off-diagonal terms of the form $\ket{b}_l^m\bra{b}_{l^\prime}^{m\pm2}$), $\exp\left[-2\gamma t\right]$  (corresponding to the off-diagonal terms of the form $\ket{b}_l^m\bra{b}_{l^\prime}^{m\pm4}$), and $\exp\left[-\frac{9}{2}\gamma t\right]$  (corresponding to the off-diagonal terms of the form $\ket{b}_l^m\bra{b}_{l^\prime}^{m\pm6}$). 
Explicit calculation of the expectation values of the quantities discussed in Eqs.~(\ref{eq:logical_op_exp_values})-(\ref{eq:operator_expt_values_in_code_space}) in Sec.~\ref{subsec:observables} under global dephasing noise leads to: 
\begin{eqnarray}
\label{eq:global_logical_3_x}
R_x&=&\text{e}^{-2\gamma t}\sin\theta\cos\phi,\\
\label{eq:global_logical_3_y}
R_y&=&\text{e}^{-\frac{1}{2}\gamma t}\sin\theta\sin\phi,\\ 
\label{eq:global_logical_3_z}
R_z&=&\frac{1}{2}\text{e}^{-\frac{9}{2}\gamma t}\left[\cos\theta+2\text{e}^{4\gamma t}\cos^2\frac{\theta}{2}-1\right],\\
\label{eq:global_csp_3}
p&=&\frac{1}{2}\Big[\text{e}^{-\frac{1}{2}\gamma t}\cos^2\frac{\theta}{2}+\text{e}^{-\frac{9}{2}\gamma t}\sin^2\frac{\theta}{2}+1\Big],\\
\label{eq:global_logical_csp_3_x}
p_x&=&\text{e}^{-\frac{5}{4}\gamma t}\sin\theta\cos\phi\cosh\frac{3\gamma t}{4}, \\
\label{eq:global_logical_csp_3_y}
p_y&=&\text{e}^{-\frac{5}{4}\gamma t}\sin\theta\sin\phi\cosh\frac{3\gamma t}{4}, \\
\label{eq:global_logical_csp_3_z}
p_z&=&\frac{1}{2}\Big[\cos\theta-\text{e}^{-\frac{9}{2}\gamma t}\sin^2\frac{\theta}{2}+\text{e}^{-\frac{1}{2}\gamma t}\cos^2\frac{\theta}{2}\Big].
\end{eqnarray}

\begin{figure*}[htb]
    \centering
    \includegraphics[width=0.8 \textwidth]{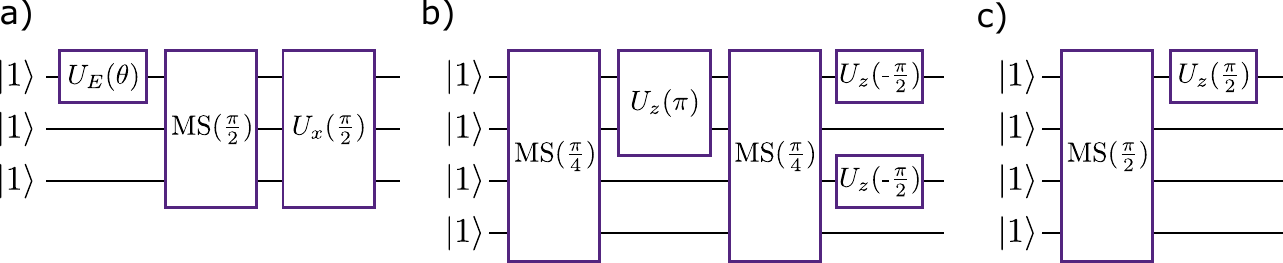}
    \caption{Circuit to prepare the a) 3-qubit $\ket{\psi}_L$ and b) 4-qubit $|0\rangle_L$ and c) 4-qubit $|+\rangle_L$ encoded states.}
    \label{fig:exp_overview}
\end{figure*}

The encoding of the logical qubit is a 3-qubit repetition code and  can be implemented by a single fully entangling MS gate with unitary $\mathrm{MS}(\pi/2)$, followed by a collective local operation collective operation $U_x(\pi/2)$~\cite{Schindler2011}. %
The individual eigenstates of the logical Pauli operators can be prepared by applying single qubit operations $U_E(\theta)=\exp(-i\theta/2 Y_1)$ on the first physical qubit before applying the MS gate. The rotation angle of $U_E(\theta)$ is $\theta \in \{0, \pi, \pi/2 \}$ to generate the $\{-1, +1, +1 \}$ logical eigenstates of the logical $\{Z_L,Z_L,X_L\}$ operators, in the following denoted as $\{-Z_L,+Z_L,+X_L\}$. The encoding circuit is shown in Fig.~\ref{fig:exp_overview}. We thus prepare the logical qubit in the +1 eigenstate of the logical X operator and the $\pm$1 eigenstates of the logical Z operators. 

In order to investigate the performance of the proposed quantities in Sec.~\ref{subsec:observables}, we let the encoded state freely evolve in time, which ideally corresponds to the implementation of the identity operation with increasing length. To get an estimate of the density matrix describing the complete system after the evolution, we perform quantum state tomography with maximum likelihood reconstruction~\cite{paris2004quantum}. We use the obtained density matrices to deduce estimates of all presented expectation values i.e. the code space stabilizers, the logical Bloch vectors and the fidelities inside the code space.

Note, that we are assessing the performance of the proposed quantities under collective dephasing noise, since this is the dominant noise source in our experimental setup. Importantly, the presented method is not limited to this type of noise and could be readily extended to other kinds of noise, like e.g. amplitude damping. One could also investigate the action of operations on logical qubits other than the identity, by e.g. performing logical randomized benchmarking~\cite{combes2017logical}, which is beyond the scope of this work.

\begin{figure*}[htb]
    \centering
    \includegraphics[width=0.8 \textwidth]{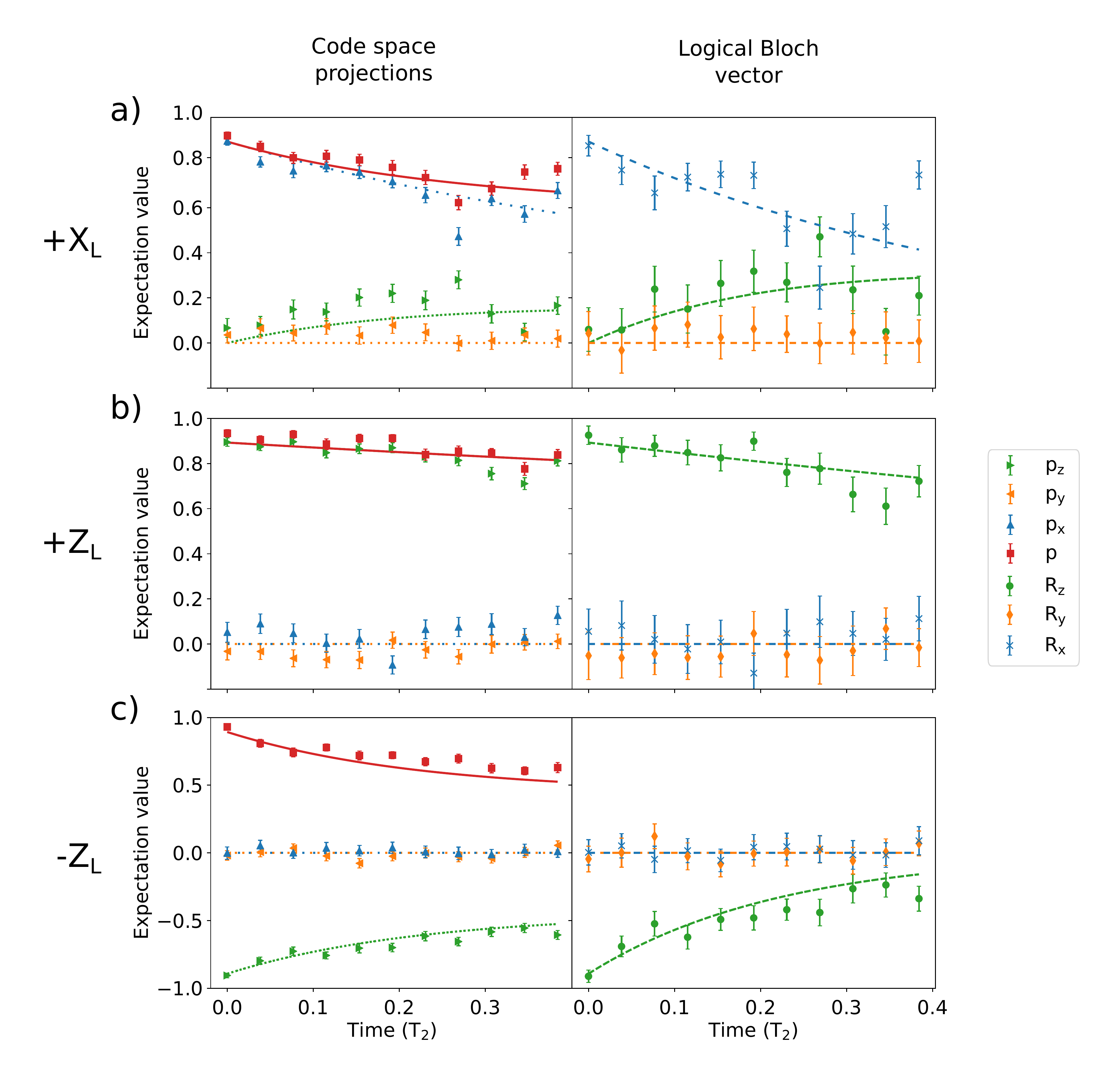}
    \caption{Expectation values of the logical Pauli operators and code space population for the 3 qubit code, initially in the a) +1 eigenstate of the logical X operator,
    b) +1 eigenstate of the logical Z operator, c) -1 eigenstate of the logical Z operator. The wait time for experimental data is given in units of $T_2=78(12)$ms and the theoretical expectation values are multiplied by a constant value of $0.89(3)$.}
    \label{fig:3qubit}
\end{figure*}

In Fig.~\ref{fig:3qubit} we present measured data of the dynamics after preparing the logical state in the $\{+1,+1,-1\}$ eigenstate of the logical $\{X_L,Z_L,Z_L\}$ operator. We estimate the coherence time of the physical qubits by performing least-squares fits of the dynamics of the individual expectation values according to Eqs.~(\ref{eq:global_logical_3_x}) - (\ref{eq:global_logical_csp_3_z}), where the experimental imperfections are modeled by multiplying the expectation value with a constant contrast factor. The mean value of all individual fit results yields an experimental coherence time $T_2=78(12)$\,ms and a contrast 0.89(3), where the error describes the standard deviation of the mean. A detailed discussion of the influence of slow drifts in the dephasing noise can be found in the appendix~\ref{sec:app_drifts}. All lines depicted in Fig.~\ref{fig:3qubit} represent the theoretical models with the mean coherence time and contrast estimated from experimental data. The relatively large standard deviation comes dominantly from laser frequency and magnetic field fluctuations in the experimental apparatus, and also from the fact that the method in its current form is not robust against state preparation and measurement (SPAM) errors.
Nevertheless, the measured data can be described very well by the theory model based on collective phase noise where the SPAM error is included as a constant contrast factor as shown in Fig.~\ref{fig:3qubit}.

Notably, in Fig.~\ref{fig:3qubit} a) the expectation value of the logical $Z_L$ operator initially vanishes but then grows with increasing storage time, as predicted by Eq.~(\ref{eq:global_logical_3_z}). This is counter-intuitive to the expectation from dephasing from physical qubits. Furthermore, this behaviour cannot be described by a quantum channel that originates from a Lindblad master equation with a time independent rate acting only on the logical qubit.
An animation of the logical qubit behavior on the Bloch sphere can be found in the online supplementary material~\cite{pal_amit_kumar_2020_4321279}.

The expectation values of the +1 and -1 eigenstate of the logical $Z_L$ operator depicted in figure~\ref{fig:3qubit} b) and c) are expected to show drastically different dynamics according to Eq.~(\ref{eq:global_logical_csp_3_z}), which is reflected in the experimental data.
Animations of the logical qubit behavior on the Bloch sphere can be found in the online supplementary material~\cite{pal_amit_kumar_2020_4321279}.

\subsubsection{Four-qubit Grassl code}
\label{sec:grassl}

Next, we consider the four-qubit QEC code used for correcting erasure noise, as proposed by Grassl \textit{et al.} \cite{PhysRevA.56.33}, defined by the stabilizers 
\begin{eqnarray} 
S_1&=&X_1X_2X_3X_4,\nonumber\\
S_2&=&Z_3Z_4,\nonumber\\\
S_3&=&Z_1Z_2. 
\end{eqnarray}
The computational basis corresponding to the logical qubit is given by $\{\ket{0}_L,\ket{1}_L\}$, with 
\begin{eqnarray}
\ket{0}_L&=&\ket{\Phi^+}\ket{\Phi^+},\;\ket{1}_L=\ket{\Phi^-}\ket{\Phi^-},
\label{eq:logical_computational_basis_4qubit}
\end{eqnarray}
where $\ket{\Phi^\pm}=\frac{1}{\sqrt{2}}(\ket{00}\pm\ket{11})$, and the logical operators are 
\begin{eqnarray}
X_L=Z_1Z_3,\,Z_L=X_1X_2,\,Y_L=-Y_1X_2Z_3.
\label{eq:logical_operator_4qubit}
\end{eqnarray}
The forms of $\{\ket{0}_L,\ket{1}_L\}$ suggest that the different magnetization values corresponding to the basis states $\ket{b}_l^m$ contributing in $\ket{\psi}_L$ are $m=4,0,-4$. This implies that the coefficients of terms of the form $\ket{b}_l^m\bra{b}_{l^\prime}^{m\pm 4}$ in the density matrix would decay as $\exp\left[-2\gamma t\right]$, while the coefficients of the terms of the form $\ket{b}_l^m\bra{b}_{l^\prime}^{m\pm 8}$ would have a time dependence given by $\exp\left[-8\gamma t\right]$. These characteristic time-decays yield decays of the expectation values (see Sec.~\ref{subsec:observables}), as given by the following equations:
\begin{eqnarray}
\label{eq:global_logical_4_x}
R_x&=&\sin\theta\cos\phi,\\
\label{eq:global_logical_4_y}
R_y&=&\econ^{-2\gamma t}\sin\theta\sin\phi,\\
\label{eq:global_logical_4_z}
R_z&=&\econ^{-2\gamma t}\cos\theta,\\
\label{eq:global_csp_4}
p&=&\frac{1}{4}\left[3+\text{e}^{-8\gamma t}+\left(\text{e}^{-8\gamma t}-1\right)\sin\theta\cos\phi\right],\\
\label{eq:global_logical_csp_4_x}
p_x&=&\frac{1}{4}\left[\text{e}^{-8\gamma t}-1+\left(\text{e}^{-8\gamma t}+3\right)\sin\theta\cos\phi\right], \\
\label{eq:global_logical_csp_4_y}
p_y&=&\text{e}^{-2\gamma t}\sin\theta\sin\phi, \\
\label{eq:global_logical_csp_4_z}
p_z&=&\text{e}^{-2\gamma t}\cos\theta.
\end{eqnarray}

\begin{figure*}[t]
    \centering
    \includegraphics[width=0.8 \textwidth]{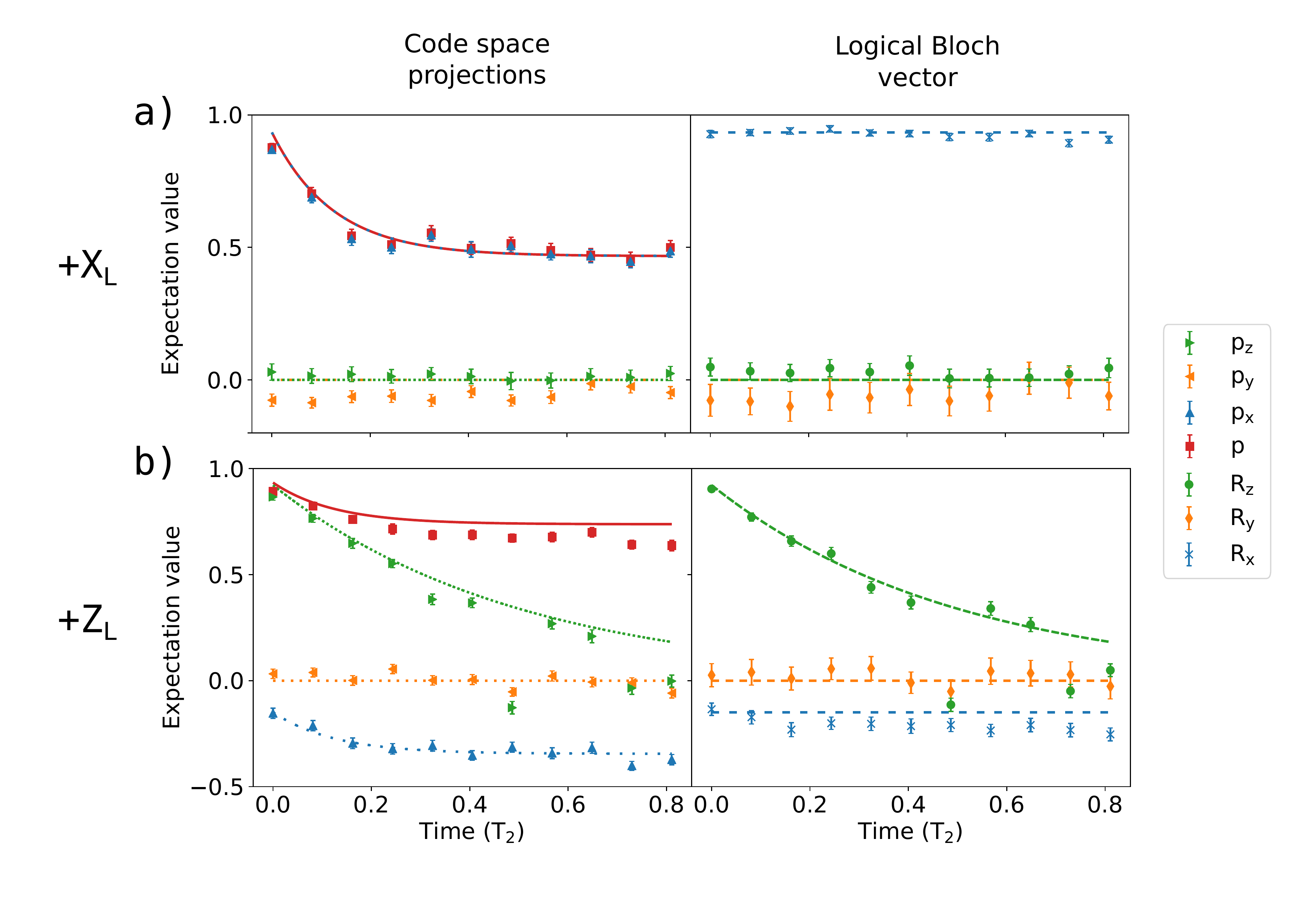}
    \caption{Expectation values of the logical Pauli operators and code space population for the 4 qubit code, initially in the a) +1 eigenstate of the logical X operator and b) the +1 eigenstate of the logical Z operator. The evolution time for experimental data is given in units of $T_2=25(5)$ms and the theoretical expectation values are multiplied by a constant value of $0.93(1)$.}
    \label{fig:4qubit}
\end{figure*}

The procedure to generate the 4-qubit Grassl code consists of two half-entangling gates MS$(\pi/4)$ 
with additional local $Z$ rotations $U_z(\theta)=\exp(-i\theta/2S_z)$, with $S_z=\sum_iZ_i$. 

For the preparation of the logical state $|0\rangle_L$ two spin echo pulses between the MS gates on qubits 1 and 2 $U_z(\pi)=\exp(-i\pi/2S_z)$, in addition to two phase correction operations $U_z(-\frac{\pi}{2})=\exp(+i\pi/4S_z)$ on qubits 1 and 3 at the end of the sequence are implemented.

The preparation of the logical state $|+\rangle_L$ has no need for spin echo pulses and hence the sequence consists of only one fully entangling gate MS$(\pi/2)$ and a single phase correction operation $U_z(\frac{\pi}{2})=\exp(-i\pi/4S_z)$ on qubit 1.

The experimental results for this four-qubit code for the +1 eigenstates of the logical X operator are shown in figure~\ref{fig:4qubit}a). Here, it is notable, that the logical X expectation value does not decay, but the population in the code space is decaying rapidly to the steady state value of 0.5.

Figure~\ref{fig:4qubit}b) shows the behavior for the +1 eigenstate of the logical Z operator. Due to miscalibrated single-qubit operations, which we discuss in detail in appendix~\ref{sec:app_calibration}, the experimentally generated eigenstate has been rotated. The theoretical description in Fig.~\ref{fig:4qubit}b) is based on a qubit in the state $\ket{\Psi}_L = \cos(\delta) \ket{0}_L + \sin(\delta) \ket{1}_L$ with $\delta=0.16$ radian.
It is notable that the expectation value of the X logical operator increases with the waiting time if the code was initially close to the +1 eigenstate of the logical Z operator. This behavior is predicted by Eq.~(\ref{eq:global_logical_csp_4_x}). Animations of the logical Bloch vectors are shown in the online supplementary material~\cite{pal_amit_kumar_2020_4321279}. The estimated coherence time is $T_2=25(5)$ms. The difference compared to the estimated 3-qubit code coherence time can be explained by the fact that the measurements were taken four months apart, where several changes to the experimental apparatus have been made in the meantime.

Note that one could work also with a variation of this code, by working with logical basis states given by
\begin{eqnarray}
\ket{0}_L&=&\ket{\Psi^+}\ket{\Psi^+},\nonumber\\
\ket{1}_L&=&\ket{\Psi^-}\ket{\Psi^-},
\label{eq:logical_computational_basis_4qubit_exp}
\end{eqnarray}
with $\ket{\Psi^\pm}=\frac{1}{\sqrt{2}}(\ket{01}\pm\ket{10})$. Note that this code is up to local single-qubit rotations equivalent to the investigated code as defined by the basis states given in Eq.~(\ref{eq:logical_computational_basis_4qubit}), however, it is expected to provide immunity against global dephasing noise. 

\section{Conclusions}
\label{sec:conclusion}
In this work, we illustrated that simple physical noise models can lead to non-trivial dynamics of logical qubits, which are not captured by usual relaxation time scales. As shown by the examples explored in this work, deviations from simple exponential decay dynamics of logical qubits are possible even in Markovian systems. However, the behavior of the encoded system can be described by the logical Pauli expectation values in conjunction with the code space population, given by the expectation value of the code-defining stabilizers.

Awareness of these effects is particularly relevant for quantum error correction protocols that protect quantum memories, where a key goal is to extend the information storage time. Here, a careful choice of logical operators, and local-unitary equivalent stabilizer operators, actually matters, and should also be taken into account when analyzing the expected performance of longer algorithms on fault-tolerant hardware. 

Extensions of the present work could include the analysis of spatial correlations which are not maximal throughout the entire register, the effect of temporal correlations, and potential generalizations of spin-echo techniques from physical qubits to logical qubits. In this regard, physically Markovian dynamics implies monotonic decay of the physical Bloch volume element~\cite{rivasRMP-2014,Lorenzo-2013}. This property can be translated to the logical level by considering the logical Bloch volume element relative to the code population. Namely, the volume element induced by the mean values $R_x^c=\Tr[\rho_c X_L]$, $R_y^c=\Tr[\rho_c Y_L]$ and $R_z^c=\Tr[\rho_c Z_L]$ for the conditional state $\rho_c=P_c \rho P_c/p_0$, which in our previous notation are nothing but $R_x^c=p_x/p_0$, $R_y^c=p_y/p_0$ and $R_z^c=p_z/p_0$. A nonmonotonic decay of this volume element certifies non-Markovian evolution at the logical level.

Furthermore, one could aim at the development of state preparation and measurement error insensitive versions of the characterization protocols used in this work. In the context of characterising logical qubits not only as quantum memories, but also logical gate operations for fault-tolerant quantum computing, first works are aiming at developing logical randomised benchmarking or gate set tomography protocols [REFS].
Finally, an interesting and open challenge concerns the derivation of effective, efficiently simulatable noise models for logical qubits. This is not only relevant for the quantum memory scenario, but also for reliable numerical predictions of the performance of logical gates or gadgets like lattice surgery, state distillation and injection techniques, which will be required for the operation of large, fault-tolerant quantum processors.

\vspace{1cm}

\textbf{Acknowledgements} We gratefully acknowledge funding by the U.S. Army Research Office (ARO) through grant no. W911NF-14-1-0103. We also acknowledge funding by the Austrian Science Fund (FWF), through the SFB BeyondC (FWF Project No. F71), by the Austrian Research Promotion Agency (FFG) contract 872766, by the EU H2020-FETFLAG-2018-03 under Grant Agreement no. 820495, and by the Office of the Director of National Intelligence (ODNI), Intelligence Advanced Research Projects Activity (IARPA), via the U.S. ARO Grant No. W911NF-16-1-0070. ll statements of fact, opinions or conclusions contained herein are those of the authors and should not be construed as representing the official views or policies of IARPA, the ODNI, or the U.S. Government.  We acknowledge support from the Institut für Quanteninformation GmbH (Innsbruck, Austria). We acknowledge financial support from the Spanish MINECO grants MINECO/FEDER Projects FIS 2017-91460-EXP,
PGC2018-099169-B-I00 FIS-2018 and from CAM/FEDER Project No. S2018/TCS- 4342 (QUITEMAD-CM). AKP acknowledges support from the National Science Center (Poland) Grant No. 2016/22/E/ST2/00559.

\vspace{5mm}
\noindent \textbf{Author contributions}. 
AKP, AE, PS, and MM wrote the manuscript and all authors provided revisions. AKP, MM, PS, and TM developed the research based on discussions with RB, AR, and MAMD. AKP and MM developed the theory. AE and PS performed the experiments and evaluated the data. AE, PS, RB, and TM contributed to the experimental setup. All authors contributed to discussions of the results and the manuscript.\\

\bibliographystyle{plainnat}

\clearpage
\onecolumngrid
\appendix
\section*{Appendix}
\label{sec:appendix}

\renewcommand{\thesubsubsection}{A.\arabic{subsection}.\arabic{subsubsection}}
\renewcommand{\thesubsection}{A.\arabic{subsection}}
\setcounter{equation}{0}
\numberwithin{equation}{section}
\setcounter{figure}{0}
\renewcommand{\theequation}{A.\arabic{equation}}
\renewcommand{\thefigure}{A.\arabic{figure}}

In this appendix, we discuss experimental imperfections due to miscalibrations and slow drifts, and we provide additional details of the action of different types of noise on encoded qubits. 

\section{Experimental imperfections due to faulty calibrations}\label{sec:app_calibration}
In the case of the 4-qubit code, as introduced in Sec.~\ref{sec:grassl}, the experiment did not result in the correct state, an eigenstate of the logical Z operator. Instead, the expectation value $R_x=\langle X_L\rangle$ shows a negative offset (and hence also $p_x=\langle X_LP_c\rangle$) already for the shortest measured evolution time, as can be seen in Fig.~\ref{fig:4qubit}b). At the same time the other expectation values $R_y=\langle Y_L\rangle$ and $R_z=\langle Z_L\rangle$ were close the theory without offset.

A possible explanation of this shifted $R_x$ value is that the single-qubit Z-operations on qubits 1 and 2, between the two MS$(\frac{\pi}{4})$ operations in Fig.~\ref{fig:exp_overview} b), were not properly calibrated. If we assume an over-rotation of the Z-operations by an angle $\delta$ as
\begin{align}
U_z(\pi)\rightarrow U_z(\pi+\delta),
\end{align}
the expectation values of the logical Bloch operators are altered to
\begin{align}
    R_x&=\langle X_L\rangle=0-\frac{1}{\sqrt{2}}\delta-\frac{\delta^2}{4}+\mathcal{O}(\delta^3)=0-\frac{1}{\sqrt{2}}\delta+\mathcal{O}(\delta^2),\nonumber\\
    R_y&=\langle Y_L\rangle=0+\mathcal{O}(\delta^3),\nonumber\\
    R_z&=\langle Z_L\rangle=1-\frac{3\delta^2}{4}+\mathcal{O}(\delta^3)=1+\mathcal{O}(\delta^2).
\end{align}
The equations above show that an over-rotation about an angle of $\delta$ has the largest effect on the offset of $R_x$, where $R_y$ and $R_z$ are insensitive to first order in $\delta$. This change in the expectation values can also be described by a superposition of the two logical states $\ket{\Psi}_L = \cos(\delta) \ket{0}_L + \sin(\delta) \ket{1}_L$ to first in $\delta$, as already introduced in the main text.

The code stabilizers are unchanged in the presence of these over-rotation to first order as can be seen in the following equations
\begin{align}
    \langle S_1\rangle&=1-\frac{3\delta^2}{4}+\mathcal{O}(\delta^3)=1+\mathcal{O}(\delta^2),\nonumber\\
    \langle S_2\rangle&=1-\frac{\delta^2}{2}+\mathcal{O}(\delta^3)=1+\mathcal{O}(\delta^2),\nonumber\\
    \langle S_3\rangle&=1-\frac{\delta^2}{2}+\mathcal{O}(\delta^3)=1+\mathcal{O}(\delta^2).
\end{align}
Thus, the dephasing dynamics of the state generated with the mis-calibrated system should predominantly follow the dynamics of a slightly rotated logical qubit. We estimate $\delta$ to be 0.16 radians from $R_x$ without any free evolution. 

For completeness, we also investigated whether an over-rotation of the MS gates (MS$(\frac{\pi}{4})\rightarrow$MS$(\frac{\pi}{4}+\delta)$), which are depicted in the circuit of Fig.~\ref{fig:exp_overview} b), can explain the rotated logical state. In this case the logical expectation value $R_x$ is independent of $\delta$ to first order and can be described as
\begin{align}
    R_x&=\langle X_L\rangle=4\delta^2+\mathcal{O}(\delta^3)>0.
\end{align}
Thus a calibration error in the angle of the MS gate can not explain the negative offset of $R_x$, as observed in Fig.~\ref{fig:4qubit}b).

\section{Experimental imperfections due to slow drifts}\label{sec:app_drifts}
In the laboratory the qubit transition frequency can change due changing magnetic fields of external devices or machines on the time scales of minutes to hours. Such slow frequency deviations can affect the quality of individual data points. One of the largest deviation can be found in the 7th data point of the expectation value $p_z$ (green triangle) in the 4-qubit encoding as illustrated in Fig.~\ref{fig:4qubit} and Fig.~\ref{fig:4qubit_noise_simulation}. At a waiting time of $\sim12$\,ms a frequency deviation of $\sim10$\,Hz or a shift in the magnetic field on the order of $\sim20$\,$\mu$G could already explain the observed imperfection. 

\begin{figure*}[t]
    \centering
    \includegraphics[width=0.3\textwidth]{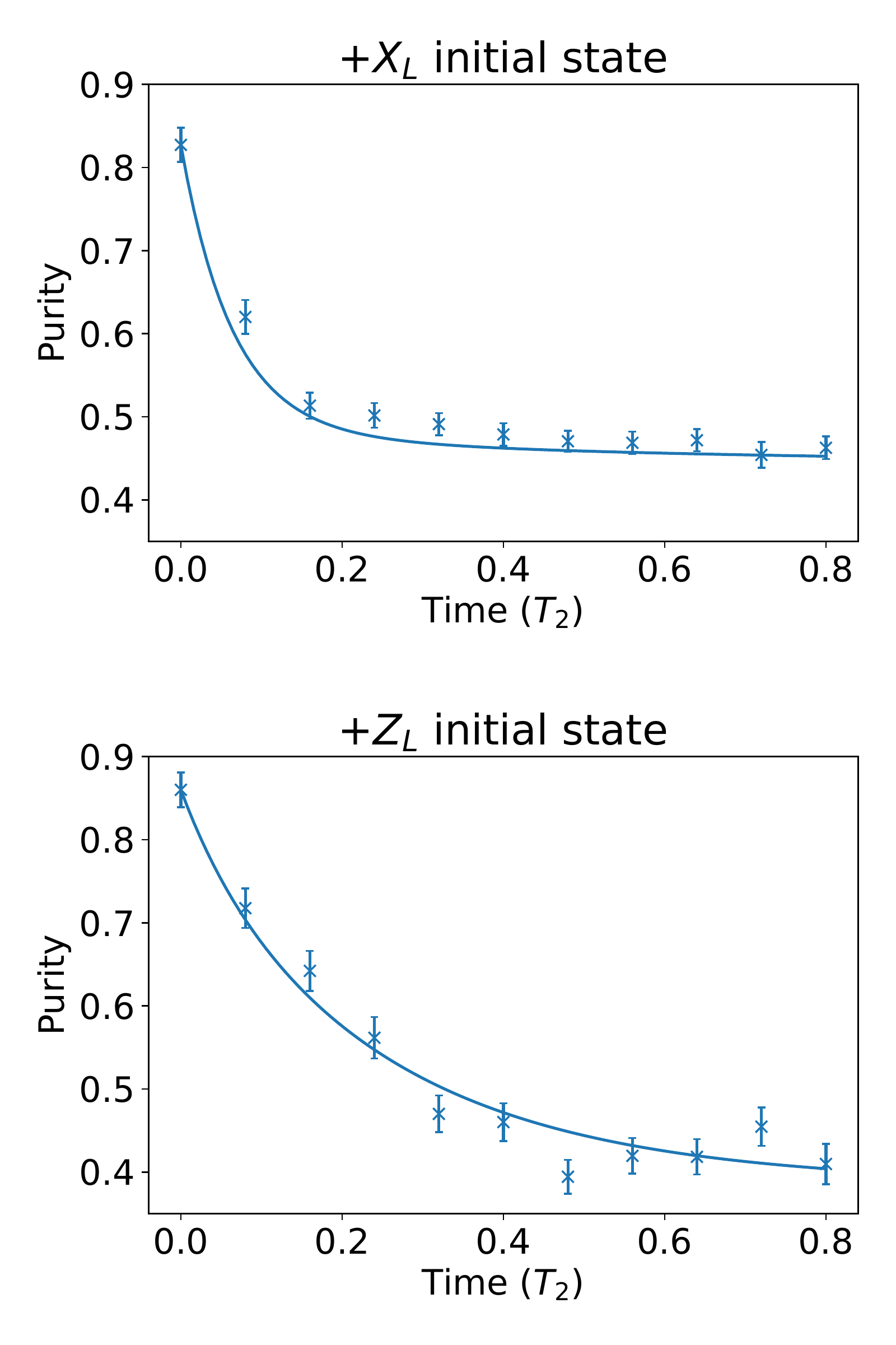}
    \caption{Time evolution of the purity Tr$(\rho^2)$ of the measured and reconstructed density matrices for the 4-qubit encoding (see Fig.~\ref{fig:4qubit}) in units of the coherence time $T_2=25(5)$\,ms. The solid lines correspond to results of a numerical simulation with global dephasing noise according to Eq.~\ref{eq:global_dephasing_noise} with coherence time $T_2=25$\,ms.}
    \label{fig:purity}
\end{figure*}

If incoherent noise were the reason for the drop in some expectation values, also the purity Tr$(\rho^2)$ of the reconstructed state would be affected. As can be observed in Fig.~\ref{fig:purity}, there is no strong deviation from the expected exponential decay in the data points. Thus we can conclude that the strong deviations in the expectation values $p_z$ are caused by coherent phase errors, probably introduced by slowly varying experimental parameters, such as e.g. the magnetic field or the laser frequency.

\section{Different types of noise affecting the encoded qubit} \label{sec:diff_noise}
As explained in the main text, we derived analytic expressions that describe the time evolution of specific observables (see Eqs.~(\ref{eq:global_logical_3_x}) - (\ref{eq:global_logical_csp_3_z})) under the action of global dephasing. Another possibility is to simulate the dissipative dynamics by numerically solving the Lindblad master equation
\begin{align}
    \dot{\rho}(t)=-\frac{i}{\hbar}\left[H(t),\rho(t)\right]+\sum_n\frac{1}{2}\left[2C_n\rho(t)C_n^\dag-\rho(t)C_n^\dag C_n-C_n^\dag C_n\rho(t)\right],
\end{align}
where $C_n=\sqrt{\gamma_n}A_n$ are collapse operators with corresponding rates $\gamma_n$.

We simulate the time evolution of the reconstructed density matrices $\rho(0)$ in the case of the three-qubit repetition code for global phase noise and amplitude damping.

\begin{itemize}
    \item \textbf{Global phase noise} can be simulated using the following collapse operator
\begin{align}
\label{eq:global_dephasing_noise}
C_1=\sqrt{\gamma/2}\left(\sigma_z\otimes\sigma_i\otimes\sigma_i+\sigma_i\otimes\sigma_z\otimes\sigma_i+\sigma_i\otimes\sigma_i\otimes\sigma_z\right),
\end{align}
where $\gamma=1/(2T_2)$ describes the dephasing rate and $T_2$ the coherence time, and $\sigma_i$ denotes the identity operator.
    \item \textbf{Amplitude damping} is simulated using the collapse operators
    \begin{align}
\label{eq:amplitude_damping_noise}
\begin{split}
C_1&=\sqrt{\gamma}(\sigma^-\otimes\sigma_i\otimes\sigma_i),\\
C_2&=\sqrt{\gamma}(\sigma_i\otimes\sigma^-\otimes\sigma_i),\\
C_3&=\sqrt{\gamma}(\sigma_i\otimes\sigma_i\otimes\sigma^-),
\end{split}
\end{align}
where $\gamma=1/(2T_1)$ describes the dephasing rate and $T_1=T_2/2$ the lifetime.
\end{itemize}
For the 4-qubit code we use the same collapse operators acting on four instead of three qubits.

\begin{figure*}[t]
    \centering
    \includegraphics[width=1.0 \textwidth]{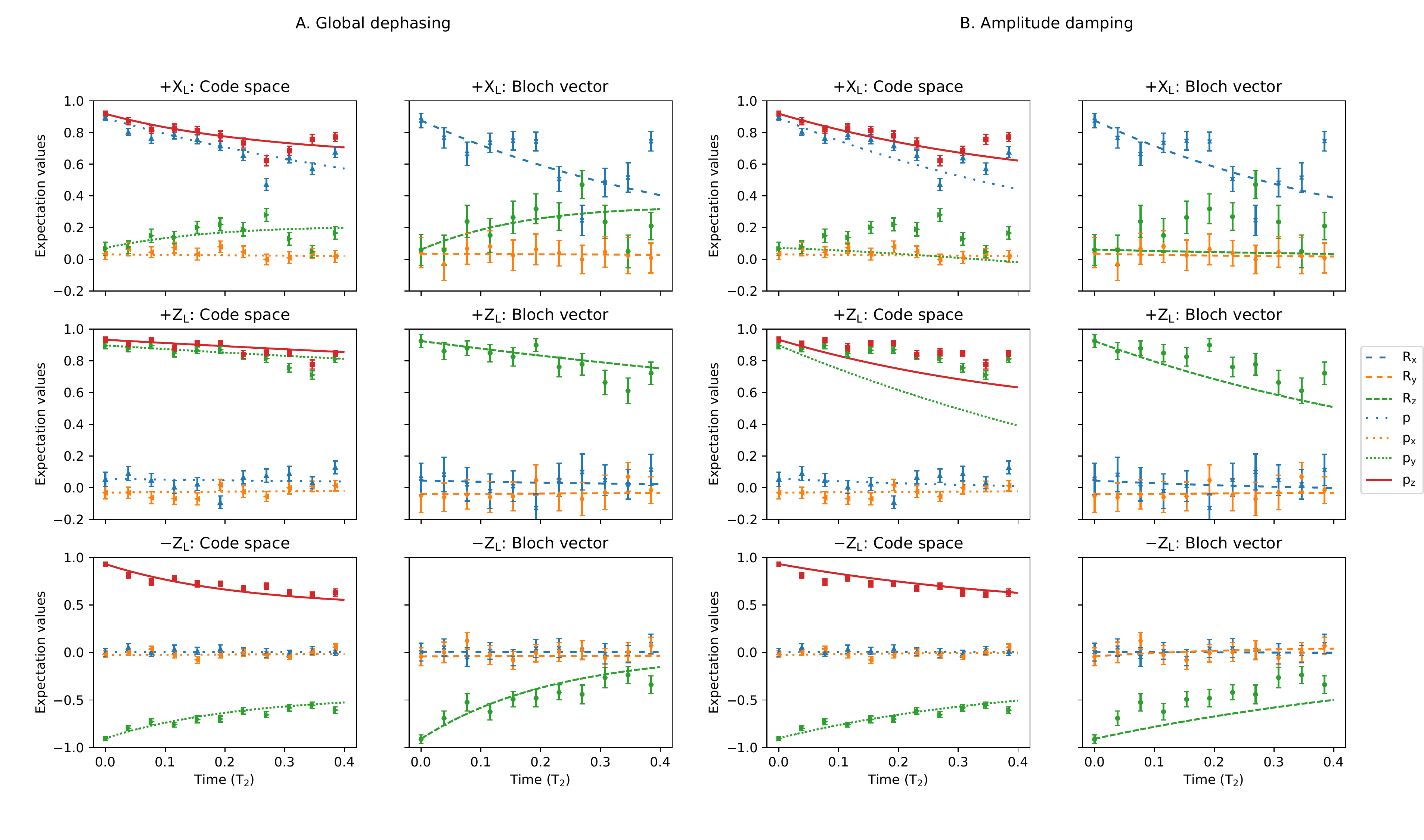}
    \caption{Time evolution simulation under the action of noise. The data points represent the experimental data for the 3-qubit encoding as shown in Fig.~\ref{fig:3qubit}. Here the lines correspond to the numerical solution of the Lindblad master equation for \textbf{A.} Global dephasing (see Eq.~\ref{eq:global_dephasing_noise}) and for \textbf{B.} Amplitude damping (see Eq.~\ref{eq:amplitude_damping_noise}) using the estimated coherence time $T_2=2T_1=78$\,ms.}
    \label{fig:3qubit_noise_simulation}
\end{figure*}

\begin{figure*}[t]
    \centering
    \includegraphics[width=1.0 \textwidth]{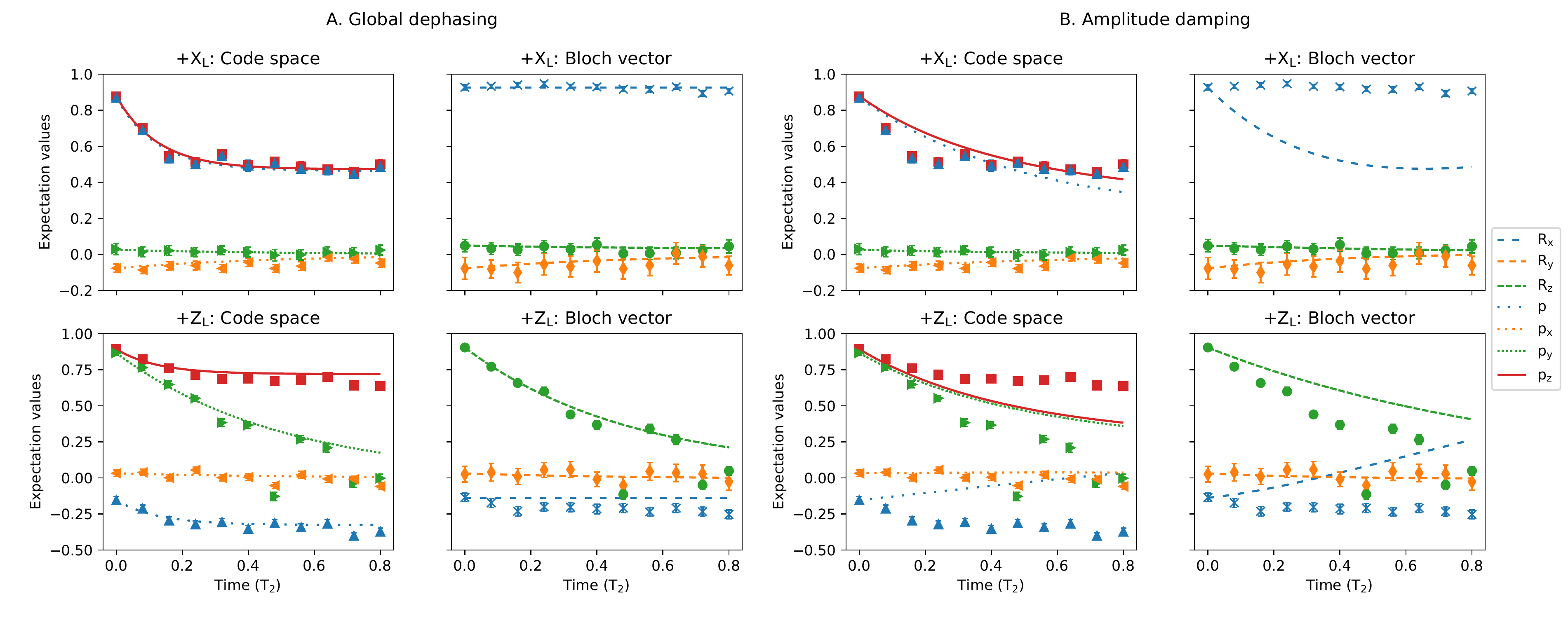}
    \caption{Time evolution simulation under the action of noise. The data points represent the experimental data for the 4-qubit encoding as shown in Fig.~\ref{fig:4qubit}. Here the lines correspond to the numerical solution of the Lindblad master equation for \textbf{A.} Global dephasing (see Eq.~\ref{eq:global_dephasing_noise}) and for \textbf{B.} Amplitude damping (see Eq.~\ref{eq:amplitude_damping_noise}) using the estimated coherence time $T_2=2T_1=25$\,ms..}
    \label{fig:4qubit_noise_simulation}
\end{figure*}

Numerical simulations of the time evolution of the encoded states under the action of different types of noise are illustrated in Fig.~\ref{fig:3qubit_noise_simulation} and in Fig.~\ref{fig:4qubit_noise_simulation}. As the initial state we use the reconstructed density matrix $\rho(T=0)$ and implement the simulation with the estimated coherence time $T_2=2T_1=78$\,ms for the 3-qubit encoding (see Fig.~\ref{fig:3qubit}) and $T_2=2T_1=25$\,ms for the 4-qubit encoding (see Fig.~\ref{fig:4qubit}) from the main text. In Fig.~\ref{fig:3qubit_noise_simulation} and ~\ref{fig:4qubit_noise_simulation} it can be very well observed, that the global phase noise simulation (A.)  properly resembles the measured dynamics, whereas the simulation using amplitude damping (B.) deviates for some expectation values clearly from the measurements.

\end{document}